\documentclass[10pt,conference]{IEEEtran}
\usepackage{epsfig,setspace,amsmath,epsf,amssymb,bm,theorem,cite,graphicx,epstopdf,algorithm,algpseudocode,float,color,mathtools,authblk}
\usepackage[table,xcdraw]{xcolor}
\usepackage{booktabs}
\usepackage{subfigure}
\usepackage{bbm}

\IEEEoverridecommandlockouts
\allowdisplaybreaks

\begin{document}

\title{Which Sensor to Observe? Timely Tracking of a Joint Markov Source with Model Predictive Control}

\author[1]{Ismail Cosandal}
\author[1]{Sennur Ulukus}
\author[2]{Nail Akar}

\affil[1]{\normalsize University of Maryland, College Park, MD, USA}
\affil[2]{\normalsize Bilkent University, Ankara, T\"{u}rkiye}
\maketitle

\begin{abstract}
In this paper, we investigate the problem of remote estimation of a discrete-time joint Markov process using multiple sensors. Each sensor observes a different component of the joint Markov process, and in each time slot, the monitor obtains a partial state value by sending a pull request to one of the sensors. The monitor chooses the sequence of sensors to observe with the goal of minimizing the mean of age of incorrect information (MAoII) by using the partial state observations obtained, which have different freshness levels. For instance, a monitor may be interested in tracking the location of an object by obtaining observations from two sensors, which observe the $x$ and $y$ coordinates of the object separately, in different time slots. The monitor, then, needs to decide which coordinate to observe in the next time slot given the history. In addition to this partial observability of the state of Markov process, there is an erasure channel with a fixed one-slot delay between each sensor and the monitor. First, we obtain a sufficient statistic, namely the \emph{belief}, representing the joint distribution of the age of incorrect information (AoII) and the current state of the observed process by using the history of all pull requests and observations. Then, we formulate the problem with a continuous state-space Markov decision problem (MDP), namely \emph{belief MDP}. To solve the problem, we propose two model predictive control (MPC) methods, namely MPC without terminal costs (MPC-WTC) and reinforcement learning MPC (RL-MPC), that have different advantages in implementation. 
\end{abstract}
\section{Introduction}
Age of incorrect information (AoII) is a joint mismatch and freshness metric that captures how long there has been a mismatch between an observed random process and its estimation at the monitor \cite{maatouk2022age}. AoII takes into account the dynamics of the source process, and thus is considered to be a semantic metric \cite{lu2022semantics, maatouk2022age}. Different from other mismatch metrics such as the mean squared error (MSE) or the binary freshness \cite{bastopcu2021,cosandal2023timely}, it penalizes the duration of incorrect estimation by increasing linearly with time and is reset to zero when the monitor correctly estimates the process. AoII is fundamentally different from other information freshness metrics including age of information (AoI) and related metrics \cite{bastopcu2020should,ayan2019value}, since these metrics reset after each status update, while AoII can be reset not necessarily only with a status update, but with an update of the estimate of the monitor, or with a state transition at the source to the estimated value at the monitor.

Partially observable Markov decision problem (POMDP) is used to formulate a Markov decision problem (MDP), where the action maker cannot directly observe the state of the problem but, from partial observations, it can estimate the likelihood of the state, which is called the \emph{belief}, and the policy is defined as a function of the belief \cite{smallwood1973optimal, spaan2012partially}. In this formulation, the state process is a Markov process, and state transition probabilities and the cost function depend only on the state and the action. In active sensing problems, on the other hand, cost or the state transition probabilities may depend on the belief, thus they do not fit the POMDP definition directly, and several formulations that extend POMDP have been proposed in \cite{araya2010pomdp, spaan2015decision, satsangi2018exploiting}. Alternatively, these extensions of POMDP and POMDP itself can be expressed with an equivalent continuous MDP where states are the distribution of unobserved states, and this formulation is called \emph{belief-MDP}.

In the freshness metrics literature, partial observability and correlated sensors are studied in different settings. In \cite{he2018minimizing, tripathi2022optimizing, zancanaro2023modeling}, there is a correlation between the processes observed by different sensors, thus getting an update from a sensor may contain partially fresh information about other processes, e.g., multiple cameras may surveil an overlapping area \cite{he2018minimizing}. Alternatively, in \cite{chen2024improving, erbayat2024age}, each sensor observes a set of processes and updates a single monitor. Therefore, samples from different sensors may include asynchronous information on the same process, and received components have different freshness levels. 

In this paper, we consider a unique case of correlated sensors, such that multiple sensors observe different components of a joint Markov process, and a monitor aims to minimize the mean of AoII (MAoII) by selecting a sensor to observe in each time slot. For example, consider an object performing a random walk on a $2D$ grid, where two sensors observe the object's $x$ and $y$ components separately, and a monitor remotely tracks the location of the object. Consider another example where a joint fire-temperature process is tracked by two sensors, one sensor observes the temperature level as low, moderate, or high, and the other sensor observes the event of a fire. In this example, we assume that the fire event only occurs if the temperature is high, thus each event is not a Markov process individually, but the joint process can be assumed to be a Markov process. These example applications will be described and investigated in more detail in Section \ref{sec:num}. 

In addition to the partial observability, we also assume a fixed delay and an erasure probability on the channel from the sensors to the monitor. Under these assumptions, the monitor can never know the exact state of the process, and hence the AoII value of its estimation. In each time slot, the monitor estimates the process using the maximum a-posteriori probability (MAP) rule, thus the AoII depends on both the source and the estimation processes. This relation disallows us to use the POMDP formulation directly. However, it is possible to obtain a \emph{belief} that corresponds to the joint distribution of the state and AoII \cite{cosandal2024joint}, and formulate the problem as a belief MDP. 

Since the belief space contains uncountably many elements, it is not possible to obtain the value function for each belief by using dynamic programming from the belief-MDP representation. Instead, model predictive control (MPC) is one of the methods used to solve belief-MDP \cite{sehr2018performance, huang2024learning, ulfsjoo2022integrating, esfahani2021reinforcement}. Starting from an initial belief state, MPC evaluates all possible outcomes for a finite steps and adds terminal costs of states in the final step \cite{bertsekas2012dynamic}. Then, from all calculated trajectories, MPC obtains the action sequences that minimize the expected cost, and applies the first action from the sequence. 

In this paper, we consider two approaches for terminal costs. The first approach is MPC without terminal cost (MPC-WTC) which considers all terminal costs to be zero, and the first action of the path which minimizes the expected cost on a finite horizon is applied \cite{boccia2014stability,grune2012nmpc}. The main advantage of this method is ease of implementation, and that it does not require offline learning. However, for a good infinite approximation, a long horizon might be required. The second one is reinforcement learning MPC (RL-MPC) where terminal costs are approximated with the aid of RL \cite{lin2023reinforcement}. It utilizes MPC-WTC to learn terminal costs in the first iteration, and increases its approximation horizon iteratively in a similar fashion to fitted Q-learning \cite{ernst2003iteratively}. That allows us to find an action sequence that minimizes the cost in a larger horizon with a low complexity compared to MPC-WTC. On the other hand, the approximation errors in previous iterations may lead to performance degradation \cite{lin2024learning}, and unlike MPC-WTC offline learning is required before implementation.

The most commonly used estimation rule in the AoII literature is the so-called \emph{martingale estimator} \cite{akar_ulukus_tcom24}, which estimates the process as the latest received status update symbol \cite{maatouk2020, maatouk2022age, chen2021minimizing, kriouile2021minimizing, kam2020age}. However, this estimation rule is not applicable to our case, since we only get a partial observation with a status update. In a recent work \cite{cosandal2024joint}, a MAP rule is utilized for the estimation that allows the monitor to update its estimation with the most likely state and, it is shown that the MAP rule is superior to the martingale estimator for AoII minimization.

The contribution of the paper can be summarized as follows:
\begin{itemize}
    \item We investigate a unique age minimization problem where a joint Markov process is observed partially, and these updates have different freshness levels.
    \item We adapt the method for calculating the distribution of age and state in \cite{cosandal2024joint} for partial observations, and formulate the problem as a belief-MDP.
    \item  To solve the belief-MDP problem, we propose two MPC methods, the first one, MPC-WTC, does not require any learning before its runtime, and the second one, RL-MPC, is able to approximate the expected cost of the problem for a longer horizon with low complexity.
\end{itemize}

\begin{figure}[t]
    \centering
    \includegraphics[width=0.95\linewidth]{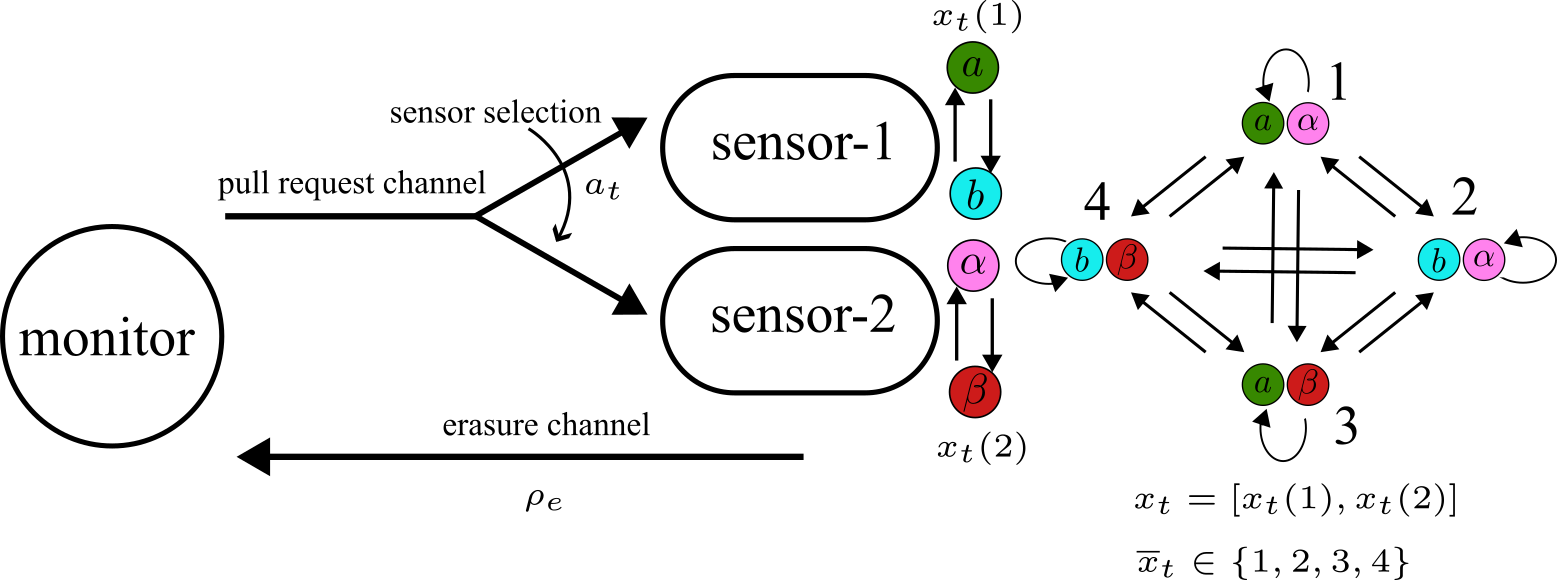}
    \caption{The illustration of the system model. The processes that sensors observe are individually non-Markov, but their joint process is Markov with known transition probabilities. State space of the joint process $x(t)$ is ordered as $\{(a,\alpha),(b,\alpha),(a,\beta),(b,\beta)\}$ and its order is denoted with $\overline{x}_t$.}
    \label{fig:sys}
    \vspace*{-0.4cm}
\end{figure}

\section{System Model}
We consider a time-slotted communication system with $K$ sensors, where sensor-$k$ observes a discrete process $x_t(k) \in \mathcal{X}_k$, where $\mathcal{X}_k$ is the distinct state-space with cardinality $|\mathcal{X}_k|=N_k$. Our main assumption is that these processes are not necessarily a Markov process individually, but the collection of these processes, which is denoted as
\begin{align}
    x_t=\{ x_t(1), \ x_t(2), \ \dots \ x_t(N) \}, 
\end{align}
is a discrete-time joint Markov process. We denote the state-space of the joint process with $\mathcal{X}$, but we usually denote the state by its index $\overline{x}_t\in\{1,2,\dots, N\}$ with the total number of joint states  $|\mathcal{X}|=N$. This process evolves with a transition matrix $P=\{p_{ij}\}$, where $p_{ij}$ denotes the transition probability from the state indexed as $i$ to state indexed as $j$.

There is a single monitor that aims to track the joint process $x_t$, or equivalently its index $\overline{x}_t$. In each time slot, the monitor takes an action $a_t=k$ to sample a single sensor $k$ and sends its decision with a reliable backward channel. With the \emph{generate-at-will (GAW)} principle, the chosen sensor sends the sampled process $x_t(k)$ via the forward channel. The forward channel is an erasure channel with a one-slot delay. In other words, the packet sampled at time $t$ can be lost in transmission with erasure probability $\rho_e$, or it is received at time $t+1$ with a successful transmission probability $\rho_s=(1-\rho_e)$. The described system model is illustrated in Fig.~\ref{fig:sys}. We denote the observation at the monitor at time $t$ with $o_t$ when the action $a_{t-1}=k$ is taken in a previous time slot as
\begin{align}
    o_t=\begin{cases}
        x_{t-1}(k), & \text{w.p.} \quad \rho_s, \\
        \emptyset,  & \text{w.p.} \quad \rho_e,
    \end{cases} \label{eq:obs}
\end{align}
where the empty-set $\emptyset$ denotes the lost packet and $o_t=x_{t-1}(k)$ denotes the received state from sensor-$k$. Additionally, the observation space for the action $a_t=k$ indicates all possible observations from sensor-$k$, thus it is defined as $\mathcal{O}(a=k)=\mathcal{X}_k \cup \emptyset.$

From all previous actions and received observations, the monitor can calculate the state distribution of the joint process as a vector $\bm{\pi}_t=\begin{bmatrix} \pi_t(1) & \pi_t(2) & \dots & \pi_N(t) \end{bmatrix}$, where each element $\pi_t(i)$ corresponds to the probability of being in state $i$, and which is expressed as
\begin{align}
    \pi_t(i)=\mathbb{P}(\overline{x}_t=i|o_1,a_1,\dots,a_{t-1},o_t).
\end{align}

The monitor estimates the joint process using the MAP rule,
\begin{align}
    \hat{x}_t=\arg\max \bm\pi_t, \label{eq:map}
\end{align}
and the mismatch is measured with the freshness metric AoII which progressively penalizes the incorrect estimation as the error stays. The exact calculation of the steady state distribution is given in the next section. We finalize this section by expressing the evolution of the AoII for time slot $t$ as
\begin{align}
    \text{AoII}_t=\begin{cases}
        \text{AoII}_{t-1}+1, & \hat{x}_t \neq \overline{x}_t, \\
        0, & \hat{x}_t = \overline{x}_t.
    \end{cases} \label{eq:MAoII}
\end{align}

\section{Belief-MDP Formulation}
The monitor aims to minimize the MAoII in the infinite horizon, based on its actions and the observations it receives. Since the monitor will never obtain the value of the process ${x}_t$ exactly due to communication delays, erasures and the partial observability of the state, it will never know the instantaneous value of AoII$_t$. Instead, the monitor estimates their distribution by using all observations it has received so far as a \emph{belief}. The distribution of AoII is dependent on the source state as stated in \cite{cosandal2024joint}. Thus, the joint belief is defined as,
\begin{align}
    b_t(i,\Delta)=\mathbb{P}(\overline{x}_t=i, \text{AoII}_t=\Delta | H_t), \label{eq:b}
\end{align}
where $H_t=\{o_1,a_1,\dots,a_{t-1},o_t\}$ is the history of all observations and actions until time $t$. For practical reasons, we truncate the age values to $\Delta_{\max}$ while calculating the belief.

\begin{figure}[t]
    \centering
    \includegraphics[width=0.95\linewidth]{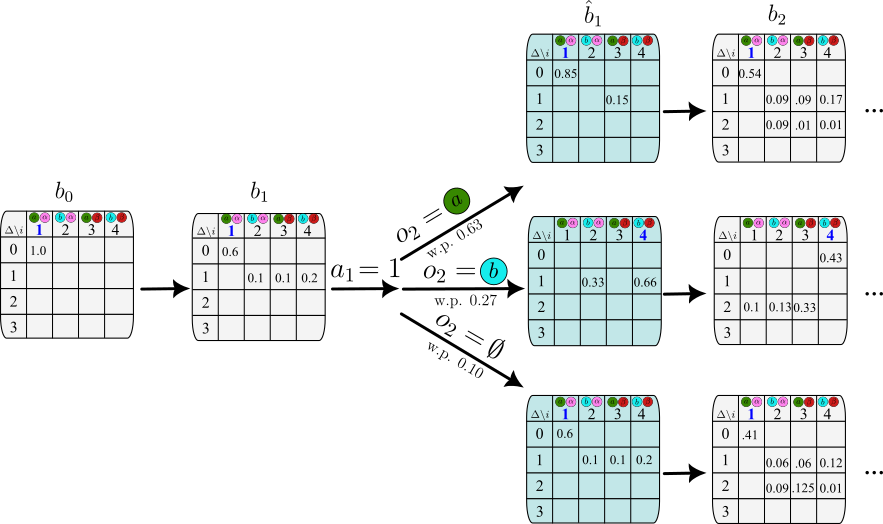}
    \caption{Evolution of the belief from an initial belief $b_0$ for the process in Fig.~\ref{fig:sys}. The first row of the transition matrix $\bm{P}$ is selected as $[0.6,0.1,0.1,0.2]$, and the remaining rows are circular-shifted one version of the previous row. After the first time step, the action $a_1=1$ is chosen, and for each possible observation $o_2\in\mathcal{O}(1)$, corresponding $\hat{b}_1$, $b_2$ and the probability of the observation is illustrated. For each case, the state with the maximum a-posteriori probability is highlighted in blue.}
    \label{fig:table}
    \vspace*{-0.4cm}
\end{figure}

We adapt the evolution of the belief in \cite{cosandal2024joint} as follows. First, assume that the monitor has the belief $b_{t-1}$ and state distribution $\bm{\pi}_{t-1}$ at the beginning of time slot $t$, and a new observation $o_t$ arrives. From \eqref{eq:obs}, if the packet is not lost during communication, the new observation includes information about $x_{t-1}$, thus the belief $b_{t-1}$ should be updated accordingly. The updated belief is $\hat{b}_{t-1}(i,\Delta)=\mathbb{P}(\overline{x}_{t-1}=i, \text{AoII}_{t-1}=\Delta|H_{t})$,  
\begin{align}
\hat{b}_{t-1}(i,\Delta)=\begin{cases}
    b_{t-1}(i,\Delta), &  o_t=\emptyset, \\
    \dfrac{ \delta_{o_t,i} b_{t-1}(i,\Delta)}{\sum_{j=1}^N b_{t-1}(j,\Delta)}, & \text{o.w.}, \label{eq:hat_b_t}
\end{cases}    
\end{align}
where $\delta_{o_t,i}$ is a indicator function that is $1$ if the observation $o_t$ includes partial observation from the state $i$. In Fig.~\ref{fig:table}, the evolution process of a belief is illustrated with an example. 

Similarly, the updated state distribution is $\hat{{\pi}}_{t-1}(i)=\mathbb{P}(\overline{x}_{t-1}=i|H_{t})$, 
\begin{align}
\hat{\pi}_{t-1}(i)=\begin{cases}
    \hat{\pi}_{t-1}(i), &  o_t=\emptyset, \\
    \dfrac{ \delta_{o_t,i} \pi_{t-1}(i)}{\sum_{j=1}^N \pi_{t-1}(j)}, & \text{o.w.} \label{eq:hat_pi_t}
\end{cases}    
\end{align}
Then, the monitor updates its estimation as $\hat{x}=\arg\max  \bm{\pi}_t$ by obtaining state distribution for time $t$ as $\bm{\pi}_t=\hat{\bm{{\pi}}}_{t-1} \bm{P}$.

Finally, the belief for time slot $t$ can be calculated using the updated belief, and the source dynamics as follows,
\begin{align}
\!\!\!\!b_t(i,\Delta)=\begin{cases}
        \max  \bm{\pi}_t, &  i=\hat{X}_t, \Delta=0, \\
        \sum_{m=1}^{N}\hat{b}_{t-1}(m,\Delta\!-\!1)p_{mi}, &  i\neq\hat{X}_t,  \Delta>0,\\
        0, & \text{o.w.}
    \end{cases} \!\label{eq:b_t}
\end{align}
We notice that $\bm{\pi}_t$ can be expressed as a function of $b_t$ since $\pi_t(k)=\sum_{\Delta=0}^{\Delta_{\max}}b_t(k,\Delta)$. It is well-known that belief $b_t$ contains all necessary information about history $H_t$ regarding decision making, and thus, it is a sufficient statistic \cite{smallwood1973optimal}. We can justify this facts using results from \eqref{eq:hat_b_t} and \eqref{eq:b_t} to obatin,
\begin{align}
    \mathbb{P}(b_{t+1}|H_{t+1})=\mathbb{P}(b_{t+1}|o_{t+1},{b}_{t})=\mathbb{P}(b_{t+1}|\hat{b}_{t}).
\end{align}

We additionally denote the evolution of $b_{t}$ from a belief $b_{t-1}=b$ and an observation $o_t=o$ as
\begin{align}
b^o=\{b_{t}|b_{t-1}=b,o_{t}=o\}. \label{eq:b_o}
\end{align}
Then, transition probabilities between belief $b_{t-1}$ and $b_{t}$ can be denoted with $T(b_t,a_{t-1},b_{t-1})$, and for $b_{t}=b_{t-1}^{o_t}$ it is equal to the probability that $o_t$ is observed. If the transmission is successful, this probability is equal to $\rho_s$ times the likelihood of the observed state in $b_{t-1}$ which can be calculated as
\begin{align}
T(b_t^{o_t\neq\emptyset},a_{t-1},b_{t-1})&=\mathbb{P}(o_t\neq\emptyset|b_{t-1},a_{t-1})\\
    &=\rho_s\sum_{i=1}^N \sum_{\Delta=0}^{\Delta_{\max}}\delta_{o_t,i} b_{t-1}(i,\Delta), \label{eq:tprob_s}
\end{align}
and the transition to $b_t^{o_t=\emptyset}$ occurs in case of an erasure with probability $\rho_e$, regardless of the action, expressed as
\begin{align}T(b_t^{o_t=\emptyset},a_{t-1},b_{t-1})&=\rho_e. \label{eq:tprob_e}
\end{align}

Finally, we can formulate the problem as belief-MDP with the tuple $\left(\mathcal{B},\mathcal{A},T(b',a,b),r(b)\right)$: In this formulation, we consider the unobserved states as $s=\{\overline{x},\Delta\}$ with a state space of $\mathcal{S}= \{1,\dots,N\} \times \{0,\dots,\Delta_{\max}\}$. Therefore, the belief space is $\mathcal{B}=[0,1]^{ N(\Delta_{\max}+1)}$, and the belief of the problem, $b \in \mathcal{B}$ is defined in \eqref{eq:b}. The action space of the problem is defined as $\mathcal{A}\in \{1,\dots,K\}$. The transition probabilities between beliefs $T(b_t,a_{t-1},b_{t-1})$ are calculated with equations \eqref{eq:tprob_s} and \eqref{eq:tprob_e}. The cost of the unobserved state is $r(s=\{\overline{x},\Delta\})=\Delta$, thus \emph{expected} cost of the belief state $b$ is equal to,
\begin{align}
    r(b)&=\sum_{i=1}^N \sum_{\Delta=0}^{\Delta_{\max}}  r^{\lambda}(s=\{i,\Delta\},a_t)b_t(i,\Delta)
    \\&=\sum_{\Delta=0}^{\Delta_{\max}}\Delta\sum_{i=1}^N b_t(i,\Delta).
\end{align}

\section{Model Predictive Control}
\begin{figure}
    \centering
    \includegraphics[width=0.95\linewidth]{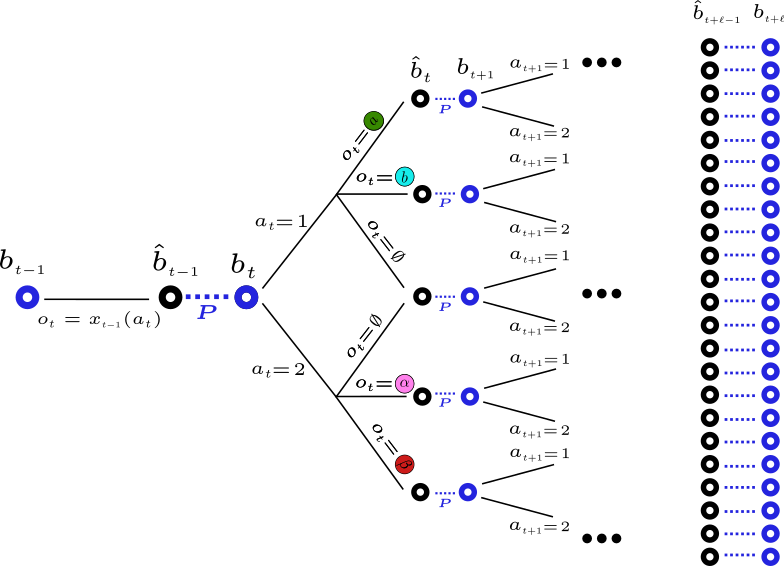}
    \caption{An $\ell$-step look-ahead table that starts a belief state and explores all possible belief states on $\ell$ steps.}
    \label{fig:rollout}
    \vspace*{-0.4cm}
\end{figure}

To find the policy that minimizes the MAoII on an infinite horizon, we utilize the MPC where the monitor searches the sequence of actions $\bm{u}^*_t=[u_t(1),\dots,u_{t}(\ell-1)]$, which minimize the expected cost by starting with a belief $b_t$, and evaluating all possible outcomes using an $\ell$-step look-ahead table. Then, the monitor applies the first action from the action sequence, i.e. $a_t^*=u^*_t(1)$. An example of the look-ahead table is given in Fig.~\ref{fig:rollout}. The action sequence $\bm{u}^*_t$ is expressed as
\begin{align}
    \underset{\substack{a_{t+m}\in \mathcal{A}, \\ m\in\{1,\dots,\ell-1\}} }{\arg\min} \mathbb{E}\left[\sum_{m=1}^\ell r(b_{t+m})+C(b_{t+\ell})\bigg|b_t,a_{m},\dots,a_{t+\ell-1}\right], \label{eq:smpc}
\end{align}
where $C(b_{t+\ell})$ is the terminal cost after step $\ell$. We propose two MPC approaches based on two terminal cost assumptions.

\subsection{MPC-WTC}
In MPC-WTC, terminal costs of beliefs are fixed to $0$. Thus, the monitor chooses the action based on the expected cost from the first $\ell$ steps. This expected value is calculated recursively by defining $R_{m,\ell}(b,a)$, the reward in next $\ell-m$ steps, belief $b$, and the action $a$, as 
\begin{align}
    \!\!\!R_{m,\ell}(b,a)=\sum_{o\in\mathcal{O}(a)} T(b^o,a,b)\left[r(b^o,a)+  R_{m+1,\ell}(b^o)\right], \label{eq:R_k}
\end{align}
where $R_{m,\ell}(b)$ is the value of belief state in $m<\ell$ step, 
\begin{align}
    R_{m,\ell}(b) = \underset{a\in\mathcal{A}}{\min} \ R_{m,\ell}(b,a),
\end{align}
and $R_{\ell,\ell}(b)=C(b)=0$ is the terminal cost which is fixed to $0$. In each time slot $t$, the monitor chooses the action as $a^*_t=\arg\min{R_{0,\ell}(b,a)}$. The recursive algorithm, which is summarized in Algorithm~\ref{alg:rec} give below, calculates $R_{m,\ell}(b)$

\begin{algorithm}[h]
    \caption{$R_{m,\ell}(b)$: The recursive algorithm to calculate the expected cost in the next $\ell-m$ steps.}\label{alg:rec}
    \begin{algorithmic}
    \For{$a \in \mathcal{A}$}          
    \For{$o \in \mathcal{O}(a)$}
    \If{$m=\ell-1$}
    \State $\sigma_{b,a,o}=C(b^o)$
    \Else
    \State $\sigma_{b,a,o}=R_{m+1,\ell}(b^o)$
    \EndIf
    \EndFor
    \EndFor
    \State \textbf{Return:} $\min_{a\in\mathcal{A}} \ \sum_{o\in\mathcal{O}(a)} T(b^o,a,b)\sigma_{b,a,o}$
    \end{algorithmic}
\end{algorithm}

\subsection{RL-MPC}
In RL-MPC, we adapt the RL-MPC method in \cite{lin2023reinforcement} that iteratively approximates terminal costs with a neural network. In the first iteration, terminal costs are $0$, similar to MPC-WTC, and the network is trained with the value of $R_{0,\ell}(b_t)$. In the following iterations, the learned parameters are used for terminal costs, extending the approximation horizon. 

For iteration $d$, we denote the terminal cost approximation as $Q(b;\theta_d)$ with the neural network parameters $\theta_d$. These parameters are trained with the loss function 
\begin{align}
    J(\theta_d)=Q(b_t;\theta_d)-R_{0,\ell}(b_t),
\end{align}
where $R_{m,\ell}(b)$ is given in \eqref{eq:R_k} except that the terminal costs are obtained from previous iteration as $R_{\ell,\ell}(b)=C(b)=Q(b;\theta_{d-1})$. The network parameters are initiated to satisfy $Q(b;\theta_d)=0$, $b\in\mathcal{B}$ for $d=1,2$.

Notice that with this method, we approximate the expected rewards in step $\ell$ in the first iteration as
\begin{align}
    Q(b;\theta_1)\approx 
    \underset{\substack{a_{t+m}\in \mathcal{A}, \\ m\in\{1,\dots,\ell-1\}} }{\min} \mathbb{E}\left[\sum_{m=1}^\ell r(b_{t+m})\bigg|b_t,a_{t},\dots,a_{t+\ell-1}\right],
\end{align}
and after iteration $d$, this is extended for the $d\ell$ steps as
\begin{align}
    Q(b;\theta_d)\approx 
    \underset{\substack{a_{t+m}\in \mathcal{A}, \\ m\in\{1,\dots,\ell-1\}} }{\min} \mathbb{E}\left[\sum_{m=1}^{d\ell} r(b_{t+m})\bigg|b_t,a_{t},\dots,a_{t+d\ell-1}\right].
\end{align}

\section{Numerical Results} \label{sec:num}
In order to investigate the performances of our proposed MPC methods, we use three belief-agnostic policies as benchmarks. The first one is \emph{random sampling}, in which the monitor chooses one of the sensors uniformly at random. The second one is \emph{round-robin}, in which the monitor chooses one of the sensors in a round-robin (circular) fashion, i.e., the action taken at time slot $t$, $a_t^{\text{rr}}$, is $a_t^{\text{rr}} = \text{mod}(a_{t-1},K)+1$, where $\text{mod}(a,b)$ is the modulo operation that returns the remainder from the division $a$ by $b$. The third one is the improved version of this method called \emph{erasure-aware round-robin}, which repeats the previous action if the message is lost in transmission, i.e., the action can expressed as $a_{t}^{\text{ea-rr}} = \text{mod}(a_{\mu(t)},K)+1$, where $\mu(t)$ is the time-stamp of the last successful transmission before time $t$.

In simulations, we assume that the initial state is known by the monitor, hence the initial AoII value is $0$. For all methods, AoII values are truncated to $\Delta_{\max}=15$, and the MAoII values are obtained over at least $10^6$ state transitions periodically starting from the initial state. For RL-MPC, a neural network with two hidden layers is used with the learning rate of $10^{-3}$, and it is trained at least $4$ iterations. We implement the following two application scenarios.

\subsection{Scenario I: Correlated Temperature-Fire-Freeze Events}
We consider a system model with three sensors. The first sensor observes the temperature of the environment. The temperature process is Markov with states of ``H'', ``M'', and ``L'' corresponding to high, moderate, and low temperatures, respectively. The remaining two sensors observe events with state space $\mathcal{X}_2=\{e_1,\overline{e}_1\}$, and $\mathcal{X}_3=\{e_2,\overline{e}_2\}$ and the event $e_1$ (resp. $e_2$) occurs only when the underlying process is in the state $x_t(1)=H$ (resp. $x_t(1)=L$). These temperature-dependent events can be considered as ``fire'' and ``freeze'', respectively. Therefore, the number of reachable joint states is reduced to $5$, and these states are sorted as $\mathcal{X}=$ $\{(H,e_1,\overline{e}_2)$, $(H,\overline{e}_1,\overline{e}_2)$, $(M,\overline{e}_1,\overline{e}_2)$, $(L,\overline{e}_1, {e}_2)$, $(L,\overline{e}_1,\overline{e}_2)\}$. Note that obtaining observations $o_{t}=e_1$, $o_{t}=$``M'', and $o_{t}=e_2$ reveals fully the joint state is $\overline{x}_{t-1}=1$, $\overline{x}_{t-1}=3$, and $\overline{x}_{t-1}=5$, respectively. Transition probabilities for the joint process are denoted with $\bm{P}_1$ as

\begin{align}
    \bm{P}_1=\begin{bmatrix}
    0.1 & 0.7 & 0.1 & 0.05 & 0.05 \\
    0.4 & 0.4 & 0.1 & 0.05 & 0.05 \\
    0.05 & 0.05 & 0.8 & 0.05 & 0.05 \\
    0.05 & 0.05 & 0.1 & 0.1 & 0.7 \\
    0.05 & 0.05 & 0.1 & 0.4 & 0.4 
    \end{bmatrix}.
\end{align}

\begin{figure}
    \centering
    \includegraphics[width=0.9\linewidth]{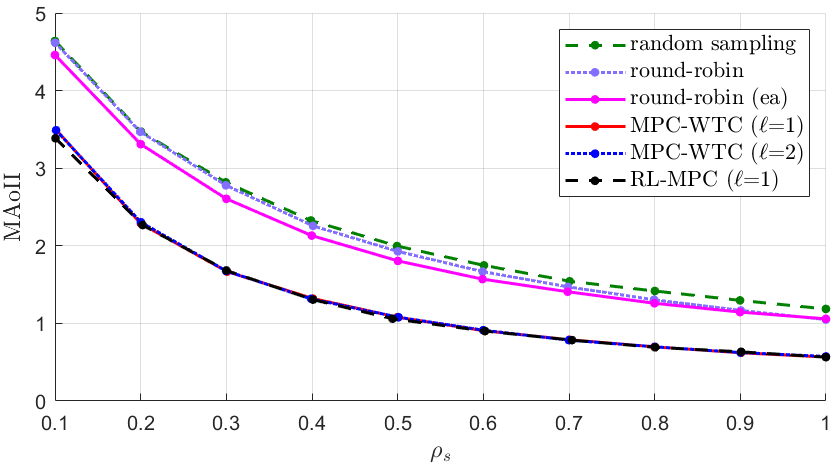}
    \caption{Comparison of methods for varying $\rho_s$.}
    \label{fig:sim_rho}
\end{figure}

Fig.~\ref{fig:sim_rho} shows the simulation results for this scenario for varying successful transmission probability $\rho_s$. We first observe that both round-robin methods work with a similar performance to random sampling with a small performance gain. Fow lower $\rho_s$, the conventional round-robin method performs similar to random sampling, and its performance convergences to the erasure-aware one with increasing $\rho_s$, as expected. On the other hand, MPC methods outperform the benchmark methods, and we observe that choosing $\ell=1$ is sufficient for this example.

\subsection{Scenario II: Random Walk on a 2D Grid}
We consider a $L_x \times L_y$ grid space, on which an object performs a random walk. We consider two sensors that are only able to observe the object's $x$ and $y$ coordinates, separately. For the random walk, we choose the probability of staying in the same state as $0.5$, moving horizontally as $0.4$ (with equal probability for right and left), and moving vertically as $0.1$ (with equal probability for down and up). For the states at corners and boundaries, these probabilities are normalized.

\begin{figure}
    \centering
    \includegraphics[width=0.9\linewidth]{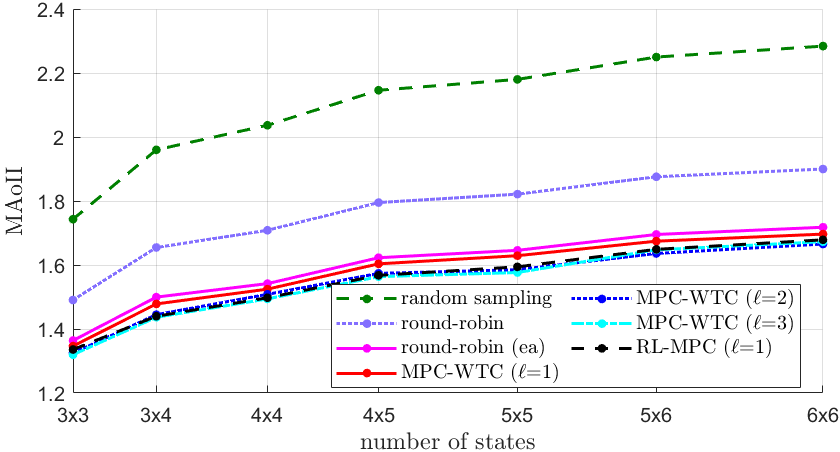}
    \caption{Comparison of methods for different grid sizes when $\rho_s=0.8$.}
    \label{fig:sim_N}
    \vspace*{-0.4cm}
\end{figure}

In Fig.~\ref{fig:sim_N}, we compare sampling methods for different state sizes $L_x \times L_y$ by fixing the erasure probability $\rho_e=0.2$ ($\rho_s=0.8$). We first observe that compared to the random sampling, round-robin methods provide performance gains, and this performance gain increases with the erasure-aware scheme. Again MPC methods outperform all benchmark methods in this scenario. We observe that MPC-WTC performs better with increasing the look-ahead step size from $\ell=1$ to $\ell=2$, and increasing it further does not improve the performance further. On the other hand, RL-MPC  captures the same performance with $\ell=1$.

\section{Conclusion}
We investigated a unique remote estimation problem in which a monitor obtains different components of a joint Markov process with different freshness levels, and it aims to minimize MAoII by choosing which sensor to observe in each time slot. First, we obtained sufficient statistics for the problem namely the belief, and formulated the problem as a belief-MDP. Then, we proposed MPC-WTC and RL-MPC methods for the problem, and compared them with benchmark methods. The main advantage of MPC-WTC is that it does not require any offline learning. On the other hand, RL-MPC has an implementation advantage that allows it to reach a similar performance to MPC-WTC with smaller step sizes. We observed that both MPC methods outperform benchmark schedules in the scenarios we investigated.

\clearpage

\bibliographystyle{unsrt}
\bibliography{bibl}

\end{document}